\documentstyle[seceq,epsf,wrapft]{ptptex}

\def\lsim{\mathrel{\mathpalette\Oversim<}}
\def\gsim{\mathrel{\mathpalette\Oversim>}}
\def\Oversim#1#2{\lower0.5ex\vbox{\baselineskip0pt\lineskip0pt%
            \lineskiplimit0pt\ialign{%
          $\mathsurround0pt #1\hfil##\hfil$\crcr#2\crcr\sim\crcr}}}
\def\RA{$\rightarrow$}
\def\Hbun{H$_2$}
\def\Hbunp{H$_2^+$}
\def\Hp{H$^+$}
\def\Hm{H$^-$}

\title{The Thermal Evolution of the Postshock Layer \\
in Pregalactic Clouds}

\author{
Hajime {\sc Susa},\footnote{E-mail:susa@rccp.tsukuba.ac.jp}
Hideya {\sc Uehara},$^{*,}$\footnote{E-mail:uehara@tap.scphys.kyoto-u.ac.jp}
Ryoichi {\sc Nishi},$^{*,}$\footnote{E-mail:nishi@tap.scphys.kyoto-u.ac.jp}\\
and Masako {\sc Yamada},$^{*,}$\footnote{E-mail:masako@tap.scphys.kyoto-u.ac.jp}
}

\inst{
Center for Computational Physics, University of Tsukuba, Tsukuba 305, Japan
\\
$^*$Department of Physics, Kyoto University, Kyoto 606-01, Japan
}

\recdate{
March 24, 1998
}

\abst{
We re-examine the thermal evolution of the postshock layer in
primordial gas clouds. Comparing the time scales, we find that the
evolutionary paths of postshock regions in primordial gas clouds can be
basically understood in terms of the diagram drawn 
in the ionization degree vs temperature plane.
The results obtained from the diagram are independent of the density
 in the case that we do not include photodissociation and photoionization.
We also argue that the diagram is not only relevant to the case of
the steady postshock flow, but also to the isochorically cooling gas. 
}

\begin{document}

\maketitle
\section{Introduction}
\label{intro}
The thermal evolution of primordial gas clouds has been investigated by
many authors. \cite{MST69,Sil77,Carl,PSS,Laha86,PD96,paperI} 
Almost all of those studies have been concerned with the formation
of various kinds of galaxies, or primordial stars. 
In those papers, galaxies are assumed to grow out of
small density perturbations present in the early universe. Because of
the coldness of the growing density perturbations, the formed clouds,
which are the progenitors of galaxies, experience strong shock heating 
at the bouncing epoch.
Shock heating is also expected in the hierarchical clustering scenario
of structure formation. In this case, shocks are expected in the
collision between two clouds which are trapped in the gravitational
potential well associated with the larger structure.

In any case, shock heating is expected in the era of galaxy
formation.
The spatial structure and the thermal evolution of the postshock layer in
primordial gas clouds were investigated by many authors. 
\cite{Izotov,StMa82a,StMa82b,MaLo86,SK,KS,AN96,YN}
In those studies the common and the most important point is the over
production of hydrogen molecules. For example, in Shapiro and Kang \cite{SK} (hereafter SK), 
the thermal evolution of steady postshock flow is investigated. 
They found that the postshock flow cools
down so fast that the recombination process cannot catch up with the cooling.
As a result, the ionization degree remains high ($y_e \sim 10^{-3}$),
even when  
the temperature has dropped below $10^4 \;{\rm K}$. 
Feeded these ``relic''
electrons, hydrogen molecules form through the processes
\begin{eqnarray}
{\rm H + e \rightarrow H^- + \gamma},\\
{\rm H + H^- \rightarrow H_2 + e}.
\end{eqnarray}
SK stressed the importance of these non-equilibrium cooling processes 
and the difficulties of describing these processes with a 
finite-difference scheme. 

SK calculated the cases corresponding to the shock velocity in the range
($50 {\rm km/s} - 300 {\rm km/s}$). In all cases, the resulting
fractions of hydrogen molecules are $\sim 10^{-3}$, irrespective of the
initial shock velocities. Ionization degrees are also independent of the
initial shock velocities for the same temperature if $T \ll 10^4$ K. 
Surprisingly, the convergence is quite good, 
though chemical equilibrium is not achieved throughout almost the entire 
 evolutionary path. 
Therefore there may be a clear explanation 
for the independence of the initial shock velocity. 
\par
In this paper we re-examine the thermal evolution of the postshock
layer in primordial gas clouds. To elucidate the convergence of
the hydrogen molecule fraction in the course of the evolution of the postshock region, 
we simply compare several important time scales (\S \ref{timesc}). 
We find that we can predict the thermal evolution of the postshock layer for any initial conditions
in a diagram of the ionization degree versus temperature.
In \S \ref{cpnum} we also test the results obtained 
in \S \ref{timesc} by comparing 
the results with those of numerical calculations for the steady
postshock flow.
In \S {\ref{conclusion}} we show that the thermal evolution of
isochorically cooling gas is very similar to that of steady postshock
flow. The remainder of the section is devoted to the further discussion and
summary.

\section{TIME SCALES}
\label{timesc}
In this section we compare the time scales which are essential to
the thermal evolution of the postshock layer in primordial gas
clouds. 
First of all, we enumerate the definition of the time scales
\begin{eqnarray}
t_{\rm cool} & \equiv & C_{p,v}\frac{\rho k T}{\mu m_p \Lambda},
 \label{eqn:tcool}\\
t_{\rm ion}  & \equiv & \frac{n_p}{k^{\rm ion} n_{\rm H} n_e} 
= \frac{1}{k^{\rm ion} n_N (1-y_e)}, \label{eqn:tion}\\
t_{\rm rec}  & \equiv & \frac{n_p}{k^{\rm rec} n_e n_p} 
= \frac{1}{k^{\rm rec} n_N y_e}, \label{eqn:trec}\\
t_{\rm dis} & \equiv & \frac{n_{\rm H_2}}
{\sum_i k^{\rm dis}_{i}  \; n_{\rm H_2} n_i} 
=\frac{1}{\sum_i k^{\rm dis}_{i}  \; n_N y_i},\label{eqn:tdis}\\
t_{\rm for} & \equiv & \frac{n_{\rm H_2}}{\sum_{ij} k^{\rm for}_{ij}\;n_i n_j} 
=\frac{y_{\rm H_2}}{\sum_{ij} k^{\rm for}_{ij}\; n_N y_i y_j}. \label{eqn:tfor}
\end{eqnarray}
Here $t_{\rm cool}, t_{\rm ion}, t_{\rm rec}, t_{\rm dis}$ and $t_{\rm for}$ denote 
the time scales of cooling, ionization, recombination, ${\rm H_2}$
dissociation, and ${\rm H_2}$ formation, respectively.
\footnote{The time scale of thermalization in the postshock layer is
shorter than the other time scales. For example, at $2\times 10^4 \;{\rm K}$
the time scales of thermalization for plasma and neutral gas are about
$10^2$ and $10^5$ times shorter than the ionization time scale, 
respectively.} 
$n_N$, $n_i$, $y_i ~(\equiv n_i / n_N)$, $k^X$, $\rho$, $\mu$ 
and  $m_p$ are the number density of nucleons,
the number density of the $i$th species, 
the fraction of {\it i}th species,
the chemical reaction rate coefficient of the ``$X$'' process,
the mass density, the mean molecular weight, and proton mass, respectively.
$\Lambda$ denotes the net cooling rate which includes radiative and 
chemical cooling/heating, assuming that the system is 
optically thin for cooling photons. This assumption is valid for 
subgalactic clouds with mass $10^6-10^8 M_\odot$ and whose densities are
smaller than $\sim 10^8\; {\rm cm^{-3}}$. If a cloud violates this
condition, the system could be optically thick for the H$_2$ cooling line photons.
Another assumption is that the primordial gas does not contain helium but only
hydrogen. This assumption affects the mean molecular weight and the
cooling rate at high temperature $(T \sim  10^5 \; {\rm K})$. However,
these are minor effects for the thermal evolution of primordial gas
clouds at lower temperature ($T \lsim 10^4$ K), 
 which we are especially interested in. 
Therefore, for simplicity we do not include helium. 
$C_{p,v}$ denotes the heat capacity of the gas cloud. 
If the gas evolves isobarically, $C_{p}$ is chosen, and if the cloud is
isochoric, $C_{v}$ is chosen. 
These quantities are expressed in terms of $\gamma$ as
\begin{eqnarray}
C_p &\equiv&\frac{\gamma}{\gamma -1}, \\
C_v &\equiv&\frac{1}{\gamma -1},
\end{eqnarray}
where $\gamma = 5/3$ for an ideal gas without molecules.
 When we are interested in a steady flow, 
the postshock layer is almost isobaric 
(see Appendix \ref{steady}).\par
The ionization time scale (Eq. (\ref{eqn:tion})) is defined as the
variation time scale of
$y_e$. Another possibility is definition in terms of $1-y_e$. However, it is
not appropriate for the problem we are now interested in, because the
magnitude of ionization degree itself is important for the formation of
hydrogen molecules. 
In any case, comparing these five time scales we can find the fastest
process without solving detailed time dependent differential equations.

We should remark that {\it all of these time 
scales are proportional to $\rho^{-1}$} 
if the cooling rate and other reaction rates are proportional to 
$\rho^2$. 
The dominant component of the cooling changes according to 
the temperature.
Below $10^4 \; {\rm K}$, the ${\rm H_2}$ line emission dominates the total
cooling of the cloud. 
In this process, the cooling rate is proportional to the number density
of the hydrogen molecules at the excited levels whose fraction is proportional
to $\rho$. \cite{HoMc}
\footnote{This treatment is limited to low density 
($n\lsim 10^4 {\rm cm^{-3}}$). 
The cooling rate is not proportional to $\rho^2$ for higher density 
(higher than the {\it critical density}).}
Hence, the cooling time scale is proportional to $\rho^{-1}$ for 
$T \lsim 10^4 \; {\rm K}$ as in Eq. ($\ref{eqn:tcool}$). 
The cooling process in $10^4 \;{\rm K} \lsim T \lsim 10^5\; {\rm K}$
is dominated by the bound-bound transition of the hydrogen atoms,
and the number of the excited hydrogen atoms is determined by 
the collisions with
other atoms. Therefore, the cooling rate is proportional to the square 
of the total density.
For higher temperature ($T\gsim 10^5\; {\rm K}$), the free-free emission 
by the collision between ions and electrons dominates the energy radiated 
away from the cloud.
In any case, the cooling rate is proportional to $\rho^2$, 
and the cooling time is proportional to $\rho^{-1}$.
\footnote{Compton cooling is effective at high redshift and high temperature. 
In this case, the cooling rate is not proportional to $\rho^2$, 
but this does not change the results significantly.}
The chemical reaction rates are proportional to $\rho^2$ if the reactions
are dominated by the collisional processes. Photoionization and H$_2$
photodissociation processes by UV photons emitted from the postshock
hot region could affect the ionization rates and the H$_2$ dissociation
rates in the postshock regions. If these radiative reactions dominate the other
reactions, total reaction rates will not be proportional to $\rho^2$, 
because photoionization and  H$_2$ photodissociation 
processes are not collisional.
Kang and Shapiro\cite{KS} investigated this
problem. The figures in their paper indicate that the effects caused by the
internally produced photons are not too large to dominate that of 
the collisional processes. Hence, as a 0th-order approximation, we
neglect the photoionization and the H$_2$ photodissociation 
processes in this paper.

As a result, the ratio of any two time scales is independent of 
the total density, and they are determined by the temperature and the chemical
compositions ($y_e$ and $y_{\rm H_2}$). As will be discussed later,
$y_{\rm H_2}$ is almost always determined by $y_e$ and $T$. Therefore,
the ratio of any two time scales just depends on $T$ and $y_e$.
In the following sections, we compare the time scales individually.
\par
\begin{figure}
   \epsfxsize=8cm
   \centerline{\epsfbox{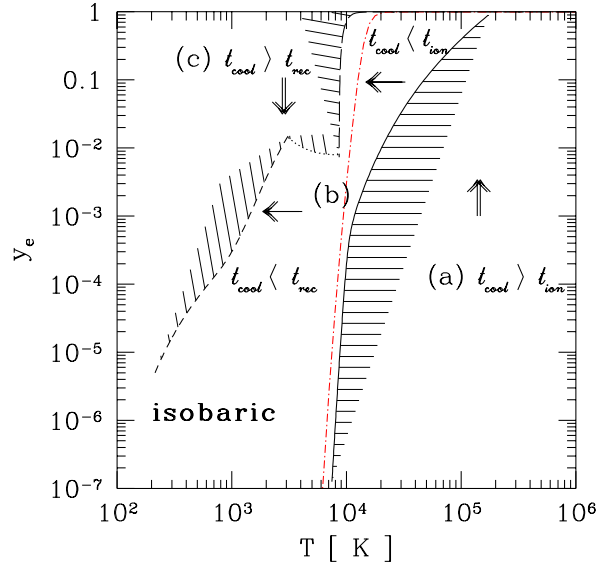}}
   \caption[dummy]{The $y_e$-$T$ plane is divided into several regions
     in which the ratio of the time scales are differ. The arrows in the plane denote the directions of the 
     evolution of the postshock layer. 
     The dot--short-dashed curve where the equality $t_{\rm rec} = t_{\rm ion}$ holds. 
         Region (a) in which the condition
     $t_{\rm ion} < t_{\rm cool}$ holds is bounded by the solid line. In
     this region, the
     system evolves upward in the $y_e$-$T$ diagram. In
     region (b), the conditions $t_{\rm cool} < t_{\rm ion}, ~t_{\rm rec}$ 
     always holds. Therefore the temperature drops down
     before the system is further ionized or recombines.
     In region (c) we have $t_{\rm rec} < t_{\rm cool}$, 
     and the system evolves downward. Region (c) is bounded by 
     the dotted line, short-dashed line and the long-dashed line.
     Those three lines denote the boundary on which the condition
     $t_{\rm cool}=t_{\rm rec}$ holds. The long-dashed line was calculated
     by assuming that the cooling rate is dominated by the hydrogen line 
     cooling. The dotted line was calculated
     assuming chemical equilibrium of the hydrogen molecules, 
     and the short-dashed line was calculated with $y_{\rm H_2} = 10^{-3}$, 
     which corresponds to the frozen value of the hydrogen molecules 
     for a shock
     velocity higher than $\sim 30\; {\rm km/s}$, see also Fig. $\ref{fig2}$.}
\label{fig1}
\end{figure}
\par
\subsection{Ionization and cooling}
\label{ion}
First, we investigate the case $t_{\rm ion} < t_{\rm rec}$, 
which corresponds to higher temperature ($T \gsim 10^4$ K), 
and we compare the time scale of ionization with that of cooling.
As mentioned previously, the ratio $t_{\rm ion}/t_{\rm cool}$ does
not depend on $\rho$, but on the ionization degree $y_e$ and temperature $T$, 
because both ionization and cooling, which are dominated by atomic
hydrogen line cooling, occur through two-body collisions. 
Hence the ratio $t_{\rm ion}/t_{\rm cool}$ is determined at each point 
on the $y_e$-$T$ plane
independently of the density.
The line along which $t_{\rm ion}/t_{\rm cool}$ is equal to unity is indicated 
 by the solid line in Fig. $\ref{fig1}$. 
If the initial conditions are given below the line, 
the system is ionized before it cools down, 
because $t_{\rm ion}/t_{\rm cool} < 1$.
Otherwise, the system cools before it is ionized. 
The expected evolutionary path is also expressed by the arrows 
in Fig. $\ref{fig1}$.
\subsection{Recombination and cooling}
\label{rec}
Next we investigate the case $t_{\rm rec} < t_{\rm ion}$, 
which corresponds to lower temperature, and examine the balance between 
the recombination process and the cooling process.
The recombination process is important for $T \lsim 10^4\; {\rm K}$ .
When $T\lsim 10^4\; {\rm K}$, the line cooling of the hydrogen molecules 
dominates the atomic hydrogen line cooling.
Therefore the ratio $t_{\rm rec}/t_{\rm cool}$ depends 
not only on the ionization degree and the temperature 
but also on the fraction of hydrogen molecules. 
In order to draw the line $t_{\rm rec}/t_{\rm cool} = 1$ in the 
$y_e$-$T$ plane,
we need information on the fraction of hydrogen molecules.
As will be discussed in the next subsection, 
the fraction of ${\rm H_2}$ is determined by the chemical
equilibrium above $\sim 10^4\; {\rm K}$. Then, it is approximately frozen 
below $\sim 4\times 10^3\; {\rm K}$  through the evolutionary course 
with cooling.
The line along which $t_{\rm rec}/t_{\rm cool} = 1$ is satisfied 
is drawn in Fig. $\ref{fig1}$.
In the region above the line, the system recombines before it cools down, and
below the line, the system cools down before the recombination proceeds.
The expected evolutionary path is also indicated in Fig. $\ref{fig1}$ by the arrows.
\subsection{Fraction of ${H_2}$}
\label{h2frac}
Here 
we investigate whether or not ${\rm H_2}$ is in chemical equilibrium.
If ${\rm H_2}$ is in chemical equilibrium (with given $T$ and $y_e$),
the cooling time is always determined by just $T$ and $y_e$.
As a result, the thermal evolution of the system is completely
determined in the $y_e$-$T$ plane.\footnote{The other components such as 
  ${\rm H_2^+}$ or ${\rm H^-}$ are always in chemical equilibrium,
  because they are very fragile.}
The fraction of hydrogen molecules is roughly determined 
by estimating $t_{\rm dis}$ 
and $t_{\rm for}$, which denote the time scale of the dissociation and 
the formation of hydrogen molecules.
If at least one of the two time scales is shorter than
the other time scales, $t_{\rm cool}$ and $t_{\rm rec}$, 
the fraction of hydrogen molecules changes toward  
the chemical equilibrium value.
Hence, the ratios $t_{\rm dis}/t_{\rm cool,rec}$ and 
$t_{\rm for}/t_{\rm cool,rec}$ should be examined. 

First of all, we comment on the important properties of $t_{\rm dis}$ and $t_{\rm for}$. 
The time scale of dissociation is independent of $y_{\rm H_2}$, 
the fraction of hydrogen molecules.
On the other hand, $t_{\rm for}$ is proportional to $y_{\rm H_2}$.
Then both the time scales given in terms of their equilibrium
values are
\begin{eqnarray}
t_{\rm dis}& = &t_{\rm dis}^{eq}, \label{eqn:diseq}\\
t_{\rm for}& = &t_{\rm for}^{eq} \frac{y_{\rm H_2}}{y_{\rm H_2}^{\it eq}},
\label{eqn:foreq}
\end{eqnarray}
where the superscript {\it eq} denotes the value at which the hydrogen molecules
are in chemical equilibrium for a given electron abundance and
temperature. Here, the electron abundance is not an equilibrium value in general. 
Therefore the relation $t_{\rm for}^{eq}=t_{\rm dis}^{eq}$ holds.

Suppose that we know the important time scale ($t_{\rm sys}$) which
characterize the evolution of the system. It could be $t_{\rm rec},
t_{\rm cool}$, and so on.
If $t_{\rm sys}$ is
independent of $y_{\rm H_2}$, we have general arguments as follows:
1) In the case $y_{\rm H_2} > y_{\rm H_2^{\it eq}}$, dissociation proceeds faster
than the formation process. 
 Hence, $t_{\rm dis}$ controls the change of $y_{\rm H_2}$. 
This is the reason we do not have to be concerned about $t_{\rm for}$.
Since $t_{\rm dis}$ is independent of $y_{\rm H_2}$,
 $y_{\rm H_2}$ converges to $y_{\rm H_2}^{\it eq}$ 
if $t_{\rm dis} < t_{\rm sys}$ is satisfied.
2) If $y_{\rm H_2} < y_{\rm H_2}^{\it eq}$,  $y_{\rm H_2}$ is controlled by the
time scale $t_{\rm for}$, because the system will try to recover the
fraction of ${\rm H_2}$ which is smaller than the equilibrium value.
The maximal value of $t_{\rm for}$ is the equilibrium value, since $t_{\rm for}$ 
is proportional to $y_{\rm H_2}$(Eq.($\ref{eqn:foreq}$)).
Therefore, if $t_{\rm sys} > t_{\rm for}^{eq}( =t_{\rm dis} )$, 
$t_{\rm sys}$ is always greater than $t_{\rm for}$.
As a result, 
$t_{\rm sys} > t_{\rm dis}$ is a sufficient condition
for the system to maintain the chemical equilibrium of hydrogen molecules.
In the following, we look into the concrete cases that the interesting time
scales are $t_{\rm rec}$ and $t_{\rm cool}$.
%
%
\par
\begin{figure}
   \epsfxsize=10cm
   \centerline{\epsfbox{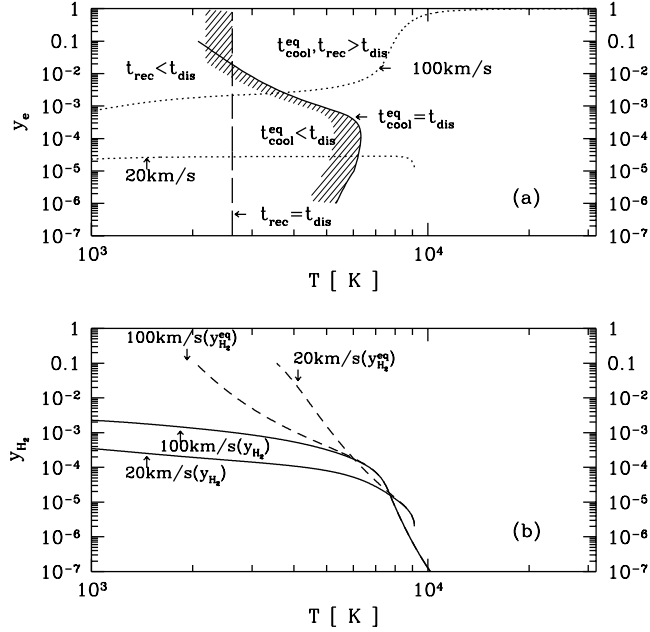}}
   \caption{
        (a) The $y_e$-$T$ plane is divided into two regions, 
        where chemical equilibrium is achieved for the fraction of 
        hydrogen molecules and where not. 
    The solid boundary of the shaded region 
    is determined by the condition $t_{\rm cool}= t_{\rm dis}$. 
        The long-dashed line denotes 
        the boundary at which $t_{\rm rec}= t_{\rm dis}$ is satisfied. 
        In the shaded region, the chemical equilibrium of 
        hydrogen molecules breaks down, 
        and the fraction $y_{\rm H_2} \equiv n_{\rm H_2}/n_{\rm H}$ 
                freezes as the temperature and/or ionization degree drops.
        The dotted lines labeled as $100\;{\rm km/s}$
        and $20\; {\rm km/s}$ 
        denote the evolutionary path of $y_e$. 
    (b) The equilibrium and non-equilibrium fraction of hydrogen molecules
    are drawn in $y_{\rm H_2}$-$T$ plane. The solid lines labeled
    $100\;{\rm km/s}(y_{\rm H_2})$ and $20\; {\rm km/s}(y_{\rm H_2})$ 
    denote the dynamically
    calculated evolutionary path of the fraction of hydrogen
    molecules.
    The short-dashed lines which are labeled as 
    $100\;{\rm km/s}(y_{\rm H_2}^{eq})$ and
    $20\; {\rm km/s}(y_{\rm H_2}^{eq})$ denote the fraction of ${\rm H_2}$
    calculated 
    assuming the chemical equilibrium of hydrogen molecules 
    with dynamically calculated electron density.
        }
\label{fig2}
\end{figure}
\subsubsection{$t_{\rm dis,for}$ vs $t_{\rm rec}$ }
First, let us investigate the chemical equilibrium of hydrogen molecules 
compared to the recombination process.
In other words, we investigate
 the magnitudes of $t_{\rm dis}/t_{\rm rec}$ and
$t_{\rm for}/t_{\rm rec}$.
As discussed above,
when we compare $t_{\rm dis}$ and $t_{ for}$ with other time scales which do not 
depend on  $y_{\rm H_2}$, we do not have to be concerned about $t_{\rm for}$.
In this case, $t_{\rm rec}$ is independent of $y_{\rm H_2}$, 
and hence we only have to examine $t_{\rm dis}/t_{\rm rec}$.
The time scale of recombination, $t_{\rm rec}$ (Eq.($\ref{eqn:trec}$)),
 is proportional to $y_e^{-1}$, and  
the time scale of dissociation is also proportional to $y_e^{-1}$,
because it is governed by collisions between ${\rm H^+}$ and ${\rm H_2}$.
This means that the ratio $t_{\rm dis}/t_{\rm rec}$ is independent of $y_e$.
It is also independent of $y_{\rm H_2}$, and hence the ratio just depends
on the temperature.
Therefore we have the critical temperature below which $t_{\rm dis}/t_{\rm rec}$ is
greater than unity (see Fig. \ref{fig2}). 
This critical temperature is $\sim 2600\; {\rm K}$.
This argument is different if the ionization degree is sufficiently
low for the ${\rm H - H_2}$ collision to dominate the  
${\rm H^+ - H_2}$ collision. 
However, within the regions we are interested ($y_e \gsim 10^{-5} -
10^{-6}$), the ionization degree is higher than this. 
The fact that the line $t_{\rm rec}=t_{\rm dis}$ is almost vertical in
Fig. \ref{fig2} 
confirms the validity of this treatment.
\par
%
%
\subsubsection{$t_{\rm dis,for}$ vs $t_{\rm cool}$ }
Next, we study the chemical equilibrium of ${\rm H_2}$ with respect to 
the cooling of the gas; i.e. we consider the ratio 
$t_{\rm dis}/t_{\rm cool}$ and $t_{\rm for}/t_{\rm cool}$.
However, we cannot use the previous argument in this case, 
because the cooling time scale depends on $y_{\rm H_2}$ for $T
\lsim 10^4\; {\rm K}$. 
Accordingly, we must investigate the individual case again. 
1) If $y_{\rm H_2} < y_{\rm H_2}^{\it eq}$, we obtain the relation $t_{\rm for} < t_{\rm dis}$. 
In this case, we do not have to consider $t_{\rm dis}$, because 
${\rm H_2}$ has a smaller fraction than its equilibrium value.
The ratio $t_{\rm for}/t_{\rm cool}$ is almost proportional to $y_{\rm H_2}^2$, 
because $t_{\rm for}$ is proportional to $y_{\rm H_2}$ and $t_{\rm cool}$ 
is  proportional to $y_{\rm H_2}^{-1}$.
Therefore the ratio reaches a maximal value at $y_{\rm H_2} = y_{\rm H_2}^{\it eq}$.
This means $t_{\rm for}/t_{\rm cool}$ is always smaller than 
 $t_{\rm for}^{\it eq}/t_{\rm cool}^{\it eq}$.
Then, if  $t_{\rm for}^{\it eq}/t_{\rm cool}^{\it eq}$ is smaller than unity,
${\rm H_2}$ is in equilibrium while the temperature drops.
2) If $y_{\rm H_2} > y_{\rm H_2}^{\it eq}$, we obtain $t_{\rm for} > t_{\rm dis}$. 
In this case we just check the ratio $t_{\rm dis}/t_{\rm cool}$, because
${\rm H_2}$ has larger fraction than its equilibrium value.
The ratio $t_{\rm dis}/t_{\rm cool}$ is proportional to $y_{\rm H_2}$. 
Therefore, the ratio does not reach a maximum at $y_{\rm H_2} = y_{\rm H_2}^{\it eq}$, 
but rather a minimum. Then we cannot conclude that $t_{\rm dis}^{eq}/t_{\rm cool}^{eq} < 1$ 
is a sufficient condition for the chemical equilibrium according to the cooling process.
However, in the region $y_{\rm H_2} > y_{\rm H_2}^{\it eq}$, the fraction of 
${\rm H_2}$ converges to $y_{\rm H_2}^{\it eq}$ as the system {\it cools},  
because the inclination of the short-dashed lines in
Fig. \ref{fig2}(b)(denoting the evolution of $y_{\rm H_2}^{eq}$)
is negative.  If the inclination is positive, $y_{\rm H_2}$ will depart 
from the
equilibrium value as the temperature drops.
In other words, if the hydrogen molecules are not in chemical
equilibrium in the region $y_{\rm H_2} > y_{\rm H_2}^{\it eq}$, 
the cooling process proceeds until the system cools down to the state
$y_{\rm H_2}=y_{\rm H_2}^{eq}$. 

In any case, if the condition $t_{\rm dis}/t_{\rm cool}^{\it eq} < 1$ is satisfied,
the relation  $y_{\rm H_2} \simeq  y_{\rm H_2}^{eq}$ always holds.
In Fig. \ref{fig2}(a) the line on which the condition 
$t_{\rm dis}/t_{\rm cool}^{\it  eq} = 1$ is satisfied is presented 
in the $y_e$-$T$ plane. The shaded region 
denotes the region in which chemical
equilibrium for hydrogen molecules with respect to the cooling or recombination process 
does not hold. When the system comes
into this region, the fraction of hydrogen molecules ``freeze''. The
fraction of hydrogen molecule becomes $\sim 10^{-3}$ 
if the initial temperature is higher than $\sim 2\times 10^4\; {\rm K}$.
This temperature corresponds to the initial shock velocity, $\sim 30\;
{\rm km/s}$, when we discuss the steady postshock flow.
For smaller shock
velocities, the initial condition for the chemical compositions is not 
reset, and the fraction of hydrogen molecule becomes a smaller value. 
\subsection{Shock diagram}
In the end of this section,
we present a brief
summary of \S \ref{timesc}.
The main results of this section are summarized in Fig. \ref{fig1}. The
evolution of a shock heated system is basically explained 
in the $y_e$-$T$ plane,
which we call the ``shock diagram''(Fig. \ref{fig1}). The evolutionary path
of the postshock layer is obtained by tracing the directions of the
arrows in the shock diagram. 
Postshock gas, which is heated up to high temperature, 
appears in region (a). 
In region (a), the shortest time scale is $t_{\rm ion}$ and 
the system evolves upward and enters region (b). 
In region (b), $t_{\rm cool}$ is the shortest, 
and therefore the system evolves leftward this time and comes into region (c). 
In region (c), $t_{\rm rec}$ is the shortest, and then  
the system evolves downward, and reenters region (b), where
 the system evolves leftward again and crosses the boundary of regions
 (b) and (c).
Finally, the system evolves nearly along the short-dashed curve, 
which is the boundary of regions (b) and (c).  
For postshock gas which is not heated up to above $2\times 10^4\; {\rm K}$,
the thermal evolution is somewhat different from the previous case.
In this case, when the system enters region (b) from region (a), 
the ionization degree is so low that the system does not enter 
region (c) at $T \simeq 10^4$ K.  
Then the ionization degree and the fraction of hydrogen molecules do
not ``forget'' the initial conditions of the chemical composition.
\subsection{Convergence of the ionization degree in the steady postshock flow}
\begin{figure}
   \epsfxsize=8cm
   \centerline{\epsfbox{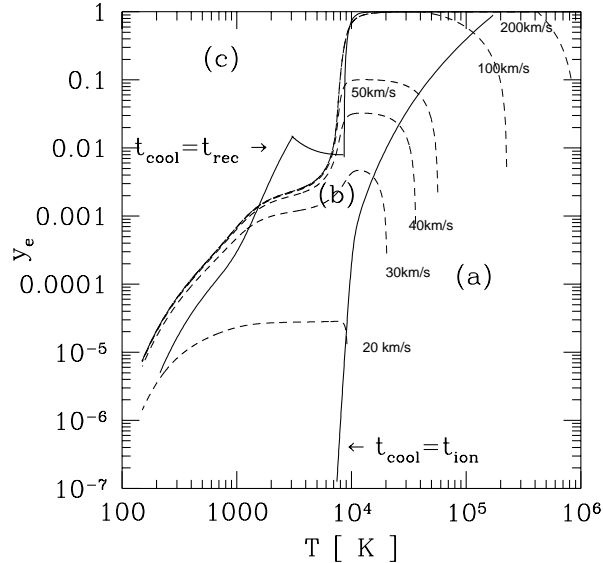}}
   \caption[dummy]{
The evolution of the postshock layer in the $y_e$-$T$ plane is presented.
The solid lines are the boundaries beyond which the direction of the
evolution is changed, and each line has the same meaning as in
Fig. $\ref{fig1}$. The short-dashed lines are the numerical results  
from the SK-type calculation. Each suffix denotes the initial shock velocity.
   }
\label{fig3}
\end{figure}

The convergences of the ionization degree and the fraction of hydrogen
molecules in the steady postshock flow are basically explained by
Figs.\ref{fig1}$\sim$\ref{fig3}.
At $T \sim 10^4\; {\rm K}$, the line 
$t_{\rm rec}=t_{\rm cool}$ becomes nearly vertical, 
since the dominant cooling process is the hydrogen atomic line cooling
which decreases by many orders below  $T \sim 2 \times 10^4\; {\rm K}$. 
If (1) the initial shock velocity is larger than $\sim 30\;{\rm km/s}$ or
(2) the initial ionization degree is larger than $\sim 7\times 10^{-3}$
and the initial temperature is larger than $T \sim 10^4\; {\rm K}$, then
the evolutionary path should hit this nearly vertical line (the long-dashed line in Fig. \ref{fig1}). 
{\it Because the line $t_{\rm rec}=t_{\rm cool}$ is nearly vertical at $T
\simeq 10^4\; {\rm K}$ and $y_e \gsim 10^{-2}$}, 
$y_e$ drops to $10^{-3} - 10^{-2}$ without changing the temperature 
and ``forgets'' the initial
conditions, though $y_e$ is not in chemical equilibrium.
The value converged to ($y_e \sim 7\times 10^{-3}$) is essentially
determined by the equation $t_{\rm rec}(y_e,T)=t_{\rm cool}^{eq}(y_e,T)$ at 
$T\simeq 8000\; {\rm K}$. 
Here, the temperature ($8000$ K) is obtained by
the equality $t_{\rm rec}/t_{\rm cool}^{Ly{\alpha}}=1$, where
$t_{\rm cool}^{Ly{\alpha}}$ is the cooling time estimated by the hydrogen
line cooling. 
The equation $t_{\rm rec}/t_{\rm cool}^{Ly{\alpha}}=1$ depends both 
on $y_e$ and on $T$ in general, but it depends just on $T$ for $y_e \ll 1$.

The convergence of ${\rm H_2}$ is also explained by Fig. \ref{fig2}.
According to Fig. \ref{fig2}, ${\rm H_2}$ is still in chemical equilibrium 
just below $T=10^4\; {\rm K}$, at which temperature the convergence of $y_e$ takes place.
As a result, systems which satisfy the previous condition (1) or (2)
experience the same state (same $y_e$, same $y_{\rm H_2}$) 
just below $T=10^4\; {\rm K}$. 
Therefore $y_{\rm H_2}$ also converges for different initial shock velocities.

\section{COMPARISON WITH NUMERICAL CALCULATIONS}
\label{cpnum}
In this section we compare the results obtained in the previous section
with numerical calculations.
The numerical calculations which we performed are the same as that in SK,
except that our calculations do not contain helium and
the cooling rate due to hydrogen molecules is  based on Hollenbach and McKee 
\cite{HoMc}\footnote{SK used the cooling rate in Lepp and Shull,
  \cite{LS} but their formulae overestimate the cooling
  rate for lower temperature, $T\lsim 10^3\; {\rm K}$.}
(see Appendices \ref{reactions} and \ref{steady}).

Figure$\ref{fig3}$ displays the evolution of $y_e$ as a function of the temperature 
behind the shock front in steady flow.
It is obvious that the expected evolutionary path in Fig. $\ref{fig1}$
agrees very well with the numerically calculated path.
We also find that there exists a critical value 
of the initial shock velocity above
which the evolution of $y_e$ below $10^4 {\rm K}$ converges.
The critical value of the shock velocity is $\sim 30\; {\rm km/s}$.
This condition for the initial shock velocity is equivalent to condition
(1) in the previous subsection.  
\par
\begin{figure}
   \epsfxsize=8cm
   \centerline{\epsfbox{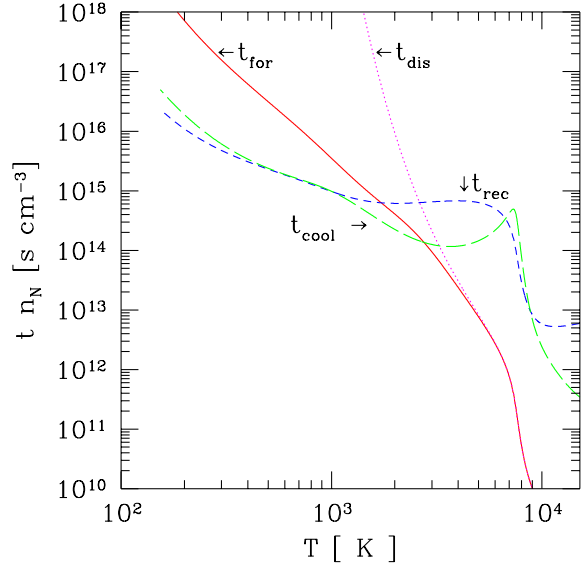}}
   \caption[dummy]{Evolution of the time scales in the steady flow
     calculation are presented. The initial shock velocity is 
     $100\;{\rm km/s}$. The solid line, dotted line, short-dashed line and
     long-dashed line represent $t_{\rm for}, t_{\rm dis}, t_{\rm rec}$
 and $t_{\rm cool}$, respectively.
     Every time scale is multiplied by density.}   
\label{fig4}
\end{figure}
The numerically calculated evolutionary path of $y_{\rm H_2}$ 
is presented in Fig. $\ref{fig2}$.
The chemical equilibrium for the hydrogen molecule holds outside of 
the shaded region in the $y_e$-$T$ plane. 
Once the system comes into the shaded region, the fraction of 
${\rm H_2}$ departs from the chemical equilibrium 
and eventually freezes to a certain value,
 because the cooling proceeds faster than
 the dissociation and formation of hydrogen molecules.
\par
The evolution of the time scales 
in numerical calculation ($v_s=100\; {\rm km/s}$) 
is presented in Fig. \ref{fig4}. 
The dissociation time scale is identical to the
formation time scale for $T\gsim\; 4000 {\rm K}$, and they are smaller
than the cooling time. For $T\lsim\; 4000 {\rm K}$, $t_{\rm dis}$ 
becomes larger than $t_{\rm cool}$, and $y_{\rm H_2}$ starts to 
depart from $y_{\rm H_2}^{eq}$. 
At a slightly lower temperature, $t_{\rm for}$ also becomes larger than $t_{\rm cool}$. 
As a result, the value of $y_{\rm H_2}$ approximately freeze for 
$T\lsim\; 4000 {\rm K}$.  
The final value of $y_{\rm H_2}$ is $10^{-3}$ for 
$v_s \gsim 30\; {\rm km/s}$, and smaller for slower shock velocities.
The frozen values degenerate for faster shock velocities, 
because the ionization degrees also degenerate 
for $v_s \gsim 30\;{\rm km/s}$.
\section{Conclusion and discussion}
\label{conclusion}
In this paper, we investigated the thermal evolution 
of the postshock layer in primordial gas clouds.
Because of the delay of the recombination process compared to 
{\it the line cooling due to the hydrogen molecules},
significant amounts of electrons ($y_e\sim 10^{-3}$)
 remain below $8000\; {\rm K}$.
Feeded the electrons, ${\rm H_2}$ also maintains high abundance  
($y_{\rm H_2}\sim 10^{-3}$)
for $T \lsim several \times 10^3\; {\rm K}$, then the cloud temperature 
drops to $T\sim 100\; {\rm K}$ quickly due to the ${\rm H_2}$ line cooling.
We have shown that both of the residual values of 
$y_e$ and $y_{\rm H_2}$ 
are universal for $v_s \gsim 30\; {\rm km/s}$ by simple comparison of time 
scales. 
The universality of the values to which $y_e$ and $y_{\rm H_2}$ converge is
explained as follows: The evolution of the system is basically 
determined by $T$, $y_e$ and $y_{\rm H_2}$. However, at high
temperature, $y_{\rm H_2}$ attains its equilibrium value, and at low
temperature, it will freeze to a value, because cooling precedes
the dissociation and the formation of hydrogen molecules.
Therefore the evolution of the system is predicted just by $T$ and $y_e$.
The convergence of $y_e$ and $y_{\rm H_2}$ occurs when the coolant
changes from hydrogen atoms to hydrogen molecules.
In fact, assuming an isobaric cooling time, the predicted evolutionary
paths agree very well with the numerically integrated paths for steady
flow.
\par
\begin{figure}
   \epsfxsize=8cm
   \centerline{\epsfbox{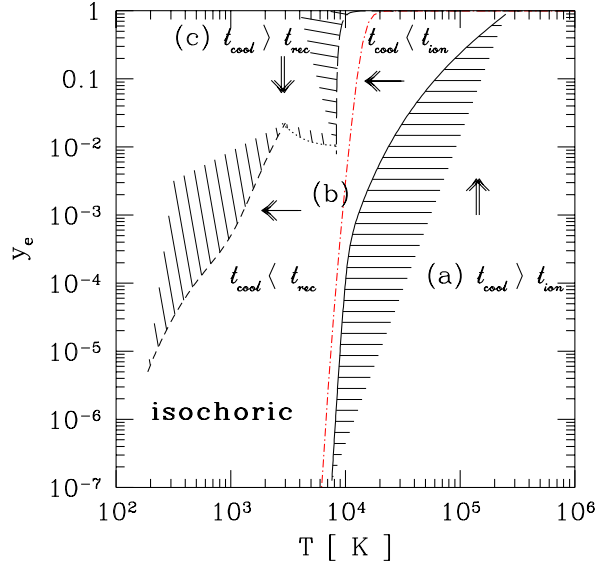}}
   \caption[dummy]{Same as Fig. \ref{fig1}, except that the cooling time is
     estimated by assuming the isochoric evolution of the cloud.} 
\label{fig5}
\end{figure}

These results are not restricted to the case of steady postshock flow,
and what is more, the shock diagram describes not only the evolution of
the postshock flow but also the primordial gas cloud which is once heated
up to $10^4\; {\rm K}$.
The argument in \S  \ref{timesc} simply depends  
on the comparison of time scales. 
When we compare the shock diagram (Fig. \ref{fig1}) with the numerically
integrated results,
the steadiness of the flow is used
just as the isobaricity of the postshock flow in Eq.
(\ref{eqn:tcool}), by choosing $C_p$.
However, even if  we choose $C_v$ in Eq. (\ref{eqn:tcool}), $t_{\rm cool}$
is not much different from the cooling time for the isobaric flow. 
We have
\begin{eqnarray}
\frac{t_{\rm cool}^{\rm isobaric}}{t_{\rm cool}^{\rm isochoric}}&=& C_p/C_v, \\
                                                        &=& \gamma.
\end{eqnarray}
Consequently, the $y_e$-$T$ diagram obtained for the isochoric case, is
very similar to Fig. \ref{fig1}. The diagram for the isochoric case is
presented in Fig. \ref{fig5}. It is obvious that the difference between
Fig. \ref{fig1} and Fig. \ref{fig5} is small. In the isochoric case, 
the final fraction of hydrogen molecules is $\sim 10^{-3}$ again, as in
the isobaric case. Hence, the intermediate case should have a similar
diagram. 
Furthermore, the shock diagrams are applied not only to
the postshock
flow but also to primordial gas clouds which are initially sufficiently
heated ($T\gsim 10^4\;{\rm K}$ ).

Hence the argument in the previous paragraph implies that 
the thermal evolution of the primordial gas cloud is determined by 
the ``shock diagram", unless the system becomes gravitationally 
unstable or some processes originating from 
external radiation fields become important. 
 The dynamical evolution of the system is roughly determined by 
the sound crossing time $t_{\rm sc}$, free fall time $t_{\rm ff}$, 
and $t_{\rm cool}$. 
In the case that $t_{\rm sc}$ is the shortest time scale, 
the evolution of the system is nearly isobaric, and 
in the case that $t_{\rm cool}$ is the shortest,
the evolution is approximately isochoric. 
In both cases, we can apply the shock diagrams on the thermal evolution
of the system.
In the third case, when $t_{\rm ff}$ is the shortest, 
the evolution is different from both of the above two cases. 
However, in a realistic situation, such as the shocks 
in dynamically collapsing primordial gas clouds,
at first the isobaric case is realized generally, as discussed by 
Yamada and Nishi.\cite{YN} 
Results of numerical simulations also imply that the shock diagrams can be
applied in this situation.  
In fact, in dynamically collapsing primordial gas clouds, 
hydrogen molecules are formed up to $y_{\rm H_2}\sim 10^{-3}$, \cite{AN96,TH} 
though the gas is not always isobaric because of a non-steady state
of the shock wave. 

The gravitational instability of sheets may change the thermal evolution 
of the primordial gas from the predicted path in the shock diagram.
The shocked and compressed layer may fragment into small pieces
with a roughly {\it local} free-fall time which is
proportional to $\rho^{-1/2}$ (Ref. \citen{ElEl}). 
In steady shock waves, Yamada and Nishi\cite{YN} discussed this problem by
comparing the time scale of fragmentation and the cooling time. 

The external UV radiation field, which exists at the early epoch of the
universe, may affect the thermal evolution of pregalactic gas.
The reaction rates of 
photoionization process and photodissociation process have different
dependence on density from the other reactions induced by collision. 
As a result, the above arguments based on Figs. \ref{fig1} and
\ref{fig5} are changed significantly.
However, the external UV radiation field
does not exist, as long as  we are interested in the first objects in
the universe.

Consequently, once the primordial gas cloud is heated above   
$10^4\; {\rm K}$
and there is no external radiation field, 
significant amounts of hydrogen molecules are formed ($y_{\rm H_2}\sim
10^{-3}$), and the temperature drops much below $10^4\; {\rm K}$, 
until some other time scales become comparable to
 the thermal and chemical time scales.

\section*{Acknowledgements}
We thank H. Sato and T. Nakamura for useful discussions. We also thank
N. Sugiyama and M. Umemura for continuous encouragement.
This work is supported in part by Research Fellowships of the Japan Society 
for the Promotion of Science for Young Scientists, No.2370(HS), 6894(HU) and
a Grant-in-Aid of Scientific Research of the Ministry of Education,
Culture, and Sports, No.08740170(RN).
\appendix
\appendix
\section{Reactions}
\label{reactions}
We include the following reactions to obtain the dynamical evolution of 
the postshock layer in \S \ref{cpnum}. The reaction rates are given 
in the listed reference in the Table \ref{tab:reaction}.
\def\RA{$\rightarrow$}
\def\Hbun{H$_2$}
\def\Hbunp{H$_2^+$}
\def\Hp{H$^+$}
\def\Hm{H$^-$}
\begin{table}
\caption{Reaction rates
\label{tab:reaction}}
\begin{center}
\begin{tabular}{cc}\hline \hline
{\bf Reactions} & {\bf References} \\ \hline
\Hp + e \RA H + $\gamma$ & Spitzer 1956 \cite{Spitz} \\
H + e \RA \Hm + $\gamma$ & de Jong 1972 \cite{dejong} \\
\Hm + H \RA \Hbun +e & Beiniek 1980 \cite{Beiniek} \\
3H \RA \Hbun + H& Palla, Salpeter \&  Stahler 1983 \cite{PSS}\\
\Hbun + H \RA 3H & SK \cite{SK} \\
2H + \Hbun \RA 2\Hbun& Palla, Salpeter \& Stahler 1983\cite{PSS}\\
2\Hbun \RA 2H + \Hbun & SK \cite{SK}\\
H + e \RA \Hp +2e & Lotz 1968 \cite{Lotz}\\
2H \RA H + \Hp + e & Palla, Salpeter \& Stahler 1983 \cite{PSS}\\
H + \Hp \RA \Hbunp + $\gamma$ & Ramaker \& Peek 1976 \cite{Ramaker}\\
\Hbunp + H \RA \Hbun + \Hp & Karpas, Anicich \& Huntress 1979 \cite{Karpas}\\
\Hbun + \Hp \RA \Hbunp + H & Prasad \& Huntress 1980 \cite{Prasad}\\
\Hm + \Hp \RA \Hbunp + e &Poularert et.al. 1978 \cite{Poulaert}\\
\Hbunp + e \RA 2H & Mitchell \& Deveau 1983 \cite{Mitchell}\\
\Hbunp + \Hm \RA H + \Hbun & Prasad \& Huntress 1980 \cite{Prasad} \\
\Hm + e \RA H + 2e &Duley 1984 \cite{Duley}\\
\Hm+ H \RA 2H + e &Izotov \& Kolensnik 1984 \cite{Izotov}\\
\Hm + \Hp \RA 2H &Duley 1984 \cite{Duley}\\ \hline
\end{tabular}
\end{center}
\end{table}

\section{Steady Shock}
\label{steady}
In this appendix, the basic equations of the steady shock are
presented. The numerical results presented in \S \ref{cpnum} are based on
the following equations.
SK solved these equations coupled with the chemical
reactions. One of the most important features of the steady shock is the
approximate isobaricity of the postshock layer.
\subsection{Basic equations}
The jump conditions at the shock front are
\begin{eqnarray}
\rho_1 v_1 &=& \rho_2 v_2, \label{eqn:jump1} \\
\rho_1 v_1^2 + p_1 &=& \rho_2 v_2^2 +p_2, \label{eqn:jump2} \\
v_1\left( \frac{\rho_1 v_1^2}{2}+\frac{\gamma_1}{\gamma_1-1}p_1 \right)
&=& u_2\left( \frac{\rho_2 v_2^2}{2}+\frac{\gamma_2}{\gamma_2-1}p_2
\right). \label{eqn:jump3}
\end{eqnarray}
These equations constitute the conservation law of the mass, momentum and energy across the shock front. The quantities whose suffixes are 1 and 2 
denote the hydrodynamical variable in 
the preshock region and the postshock region, respectively.
In case that the shock is strong, i.e. the sound velocity in the preshock
region is much smaller than the bulk velocity $u_1$, we obtain from
Eqs. (\ref{eqn:jump1})$\sim$(\ref{eqn:jump3}),
\begin{eqnarray}
\rho_2/\rho_1 &=& v_1/v_2 = 4,  \\ \label{eqn:astrong1}
kT &=& \frac{3}{16}\mu m_p v_1^2, \label{eqn:astrong2}
\end{eqnarray}
where $\mu$ and $m_p$ are the mean molecular weight of the gas and
proton mass. 
\par
Behind the shock front, the assumption that the flow is steady allows us
to write the conservation of mass, momentum and the energy as 
\begin{eqnarray}
\rho v &=& \rho_2 v_2, \\ \label{eqn:flow1}
\rho v^2 +p &=& \rho_2 v_2^2+p_2, \\ \label{eqn:flow2}
\frac{dE}{dt} &=& (\Gamma-\Lambda)+[(p+E)]\frac{1}{\rho}\frac{d\rho}{dt},
\label{eqn:flow3}
\end{eqnarray}
where $\Gamma$, $\Lambda$ and $E$ denote the heating and cooling rates, and
the internal energy per unit volume, and $d/dt$ is the Lagrange derivative.
The hydrodynamical variables with no suffix denote quantities at
a general point behind the shock front. In addition, we require the
equation of state $p=nkT$. Then the equations are closed, and we can
solve the flow equations.  
\subsection{Isobaricity of the postshock layer}
The assumption that the flow is steady indicates the approximate 
isobaricity of the postshock region (Ref.~\citen{YN}). 
From Eqs. (\ref{eqn:flow1}) and (\ref{eqn:flow2}) and 
the equation of state, $p=\rho k T /(m_p \mu )$, the relative
compression factor $x\equiv \rho/\rho_2$ is given by
\begin{eqnarray}
\frac{kT}{m_p \mu} x^2 - (1+\frac{p_2}{\rho_2 v_2^2})x+v_2^2=0. 
\end{eqnarray}
In case the shock is strong, the above equation becomes 
\begin{eqnarray}
\frac{kT}{m_p \mu} x^2 - 4 v_2^2 x +v_2^2=0. 
\end{eqnarray}
Its solution exists if the condition
\begin{eqnarray}
v_2^2(4v_2^2- \frac{kT}{m_p \mu}) \ge 0
\end{eqnarray}
 is satisfied, and this condition holds unless the heating rate is not
 too high to raise the temperature by a factor of $4/3$ in the postshock flow.

The solution is expressed as
\begin{eqnarray}
x=\frac{m_p \mu}{kT} v_2^2 \left ( 2+\sqrt{4-\frac{kT}{m_p \mu v_2^2}} \right ).
\end{eqnarray}
If we define $T_2$ as
\begin{eqnarray}
k T_2 &=& 3\mu m_p v_2^2, 
\end{eqnarray}
the pressure $p$ can be written as follows:
\begin{eqnarray}
p = \frac{2}{3} p_2 \left( 1 + \sqrt {1 - \frac{3}{4} \frac{T}{T_2}} \right).
\end{eqnarray}
Thus, $p / p_2$ can be determined by $T / T_2$ only, 
and the dependence is weak. 

After the gas has cooled down to $T \ll T_2$, 
$p$ has the asymptotic value, $4/3 p_2$, which is not much different
from $p_2$. As a result, the postshock region is almost isobaric for
a wide range of temperatures.

\def\apj{Astrophys.J.}
\def\aap{Astron and Astrophys.}
\def\mnras{Mon.Not R. Astron. Soc.}

\end{document}